\documentstyle[12pt]{article}

\begin{document}
\renewcommand{\thefootnote}{\fnsymbol{footnote}}
\begin{titlepage}
\begin{flushright}
arXiv:0710.3471 [hep-th]
\end{flushright}

\vspace{1mm}

\begin{center}
{\Large\bf Dual Actions for Born-Infeld and $Dp$-Brane Theories}
\vspace{8mm}

{\large Rong-Xin Miao}\\
\vspace{2mm} {\em Department of Physics, Nankai University, Tianjin 300071, \\
People's Republic of China}

\vspace{4mm}
{\large Yan-Gang Miao\footnote{Corresponding author.
{\em E-mail: miaoyg@nankai.edu.cn}}}\\
\vspace{2mm} {\em Department of Physics, Nankai University, Tianjin 300071, \\
People's Republic of China}

{\em The Abdus Salam International Centre for Theoretical Physics,\\
Strada Costiera 11, 34014 Trieste, Italy}

\vspace{4mm}
{\large Shao-Jie Wei}\\
\vspace{2mm} {\em Department of Physics, Nankai University, Tianjin 300071, \\
People's Republic of China}
\end{center}

\vspace{5mm}
\centerline{{\bf{Abstract}}}
\vspace{2mm}

Dual actions with respect to U(1) gauge fields for Born-Infeld and $Dp$-brane theories are reexamined.
Taking into account an additional condition,
i.e. a corollary to the field equation of the auxiliary metric, one obtains an alternative dual action that
does not involve
the infinite power series in the auxiliary metric given by ref.~\cite{s14},
but just picks out the first term from the series formally. New effective interactions of the theories are revealed.
That is, the new dual action
gives rise to an effective interaction in terms of one interaction term rather than infinite terms of different (higher) orders of
interactions physically.
However, the price paid for eliminating the infinite power series is that the new action is not quadratic but highly nonlinear
in the Hodge dual of a $(p-1)$-form field strength. This non-linearity is inevitable to the requirement the two dual actions are equivalent.


\end{titlepage}
\newpage
\renewcommand{\thefootnote}{\arabic{footnote}}
\setcounter{footnote}{0}
\setcounter{page}{2}

\section{Introduction and Summary}

It is a common point of view that the remarkable progress of string theory~\cite{s1} is the discovery of Dirichlet $p$-branes~\cite{s2}. Geometrically,
the $Dp$-branes are $(p+1)$-dimensional hypersurfaces that are embedded in
a higher dimensional spacetime. Dynamically, they are solitonic solutions to string equations that are ``branes'' on which open strings attach
with Dirichlet boundary conditions. The dynamics of $Dp$-branes is induced by the open strings and governed in general by
the action of Born-Infeld type~\cite{s3}. Recently, many different actions for generalizations of the Born-Infeld have been
proposed to describe the effective worldvolume theories of $Dp$-branes. For instance, see
refs.~\cite{s4,s5,s6,s7,s8,s9,s10,s11,s12,s13,s14,s15,s16,s17,s18}
where quite interesting are the action with quadratic abelian field strengths~\cite{s14}
and its conformal invariant development~\cite{s15}.
The motivation for introducing an auxiliary metric~\cite{s14} and restoring a conformal symmetry~\cite{s15} lies particularly in simplifying
quantization, which originates from the fact that the string action~\cite{s19} with an auxiliary worldsheet metric and conformal
invariance greatly simplifies the analysis of string theory and allows a covariant quantization~\cite{s20}. One more recent development~\cite{s18},
nevertheless,
depends on the introduction of two independent auxiliary metrics and works various new actions some of which possess
a so-called double conformal invariance.

In this paper we reexamine dual actions with respect to U(1) gauge fields for Born-Infeld and $Dp$-brane theories.
Our main point is to note that
one can eliminate the auxiliary metric and that this has some advantages we shall show later, which is contrary to that of ref.~\cite{s14}
where one eliminates the field strength instead then one gets the dual version.
In ref.~\cite{s14},
the dual action is quadratic in a new two-form field strength that can be solved in terms of the Hodge dual
of a $(p-1)$-form field strength, and involves an infinite power series in the
auxiliary metric.
This infinite power series includes in fact infinite different (higher) orders of interactions of the auxiliary metric and other fields.
The quadratic form originates from its mother action that is quadratic in the abelian field strength.
Here we give an alternative dual action in which this infinite power series somehow does {\em not} appear but just its {\em first term} remains
when we take into account
an additional condition that was unnoticed before, i.e. one of corollaries to the field equation of the auxiliary metric.
Besides its formalism with finite terms, physically this type of actions gives an effective interaction in which
infinite terms of different (higher) orders of interactions are replaced by only the lowest order of the interaction
of the auxiliary metric and other fields.
However, the price paid
is that the new dual action is highly nonlinear in
the new two-form field strength which is, as already mentioned, the Hodge dual
of a $(p-1)$-form field strength. It is the non-linearity that the two dual actions are equivalent classically.
The reason lies in the difference of the two actions that is an infinite power series of higher orders of interactions
of the auxiliary metric and other fields. If
there were no non-linearity, the two actions would not be equivalent because the infinite interacting terms could never be equal to a surface term.
In a sense, our result uncovers a new phenomenon existed in classically equivalent theories
of the BIons and $Dp$-branes, that is, the interaction expressed by an infinite power series with an independent auxiliary metric is effective to
the one that picks out only the first term from the series but with a constrained auxiliary metric.

\section{Dual Actions for the Born-Infeld}

For the sake of convenience to compare our
result with that of ref.~\cite{s14}, we use the same notation as that adopted by that reference. Let us start with the mother action
for the Born-Infeld theory in $p+1$ spacetime dimensions proposed in
ref.~\cite{s14},\footnote{For simplicity but without losing generality, let the parameter $\Lambda$ be unit in this note.}
\begin{equation}
S=-\frac{T_{p}}{4}\int d^{p+1}x(-g)^{\frac{1}{4}}(-{\gamma})^{\frac{1}{4}}\left[{\gamma}^{{\mu}{\nu}}\left(g_{{\mu}{\nu}}-g^{{\rho}{\sigma}}
F_{{\mu}{\rho}}F_{{\sigma}{\nu}}\right)-\left(p-3\right)
\right],
\end{equation}
which is quadratic in the gauge field strength $F_{{\mu}{\nu}}$ as mentioned above. Various symbols stand for as follows:
\begin{equation}
F_{{\mu}{\nu}}={\partial}_{\mu}A_{\nu}-{\partial}_{\nu}A_{\mu}
\end{equation}
is the field strength of an abelian gauge field $A_{\mu}$, some Greek lowercase letters, for example,
${\mu},{\nu},{\rho},{\sigma}$, running over $0,1,\cdots,p$, are used as spacetime indices and $g_{{\mu}{\nu}}$ is the spacetime metric.
${\gamma}_{{\mu}{\nu}}$ is the auxiliary metric that is introduced in order to rewrite the Born-Infeld action as the quadratic form in $F_{{\mu}{\nu}}$.
Different from the case occurred earlier~\cite{s13}, here the auxiliary metric is symmetric.
Moreover, $g^{{\mu}{\nu}}$ and
${\gamma}^{{\mu}{\nu}}$ mean the inverse of $g_{{\mu}{\nu}}$ and ${\gamma}_{{\mu}{\nu}}$, respectively, and
$g\equiv {\rm det}(g_{{\mu}{\nu}})$, ${\gamma} \equiv {\rm det}({\gamma}_{{\mu}{\nu}})$.

In order to deduce the dual form of eq.~(1) with respect to the gauge field $A_{\mu}$, we impose a Lagrange multiplier term upon the mother action and
thus construct such an action
\begin{eqnarray}
S^{\prime}&= &-\frac{T_{p}}{4}\int d^{p+1}x\left\{(-g)^{\frac{1}{4}}(-{\gamma})^{\frac{1}{4}}\left[{\gamma}^{{\mu}{\nu}}
\left(g_{{\mu}{\nu}}-g^{{\rho}{\sigma}}
F_{{\mu}{\rho}}F_{{\sigma}{\nu}}\right)-\left(p-3\right)\right]\right.\nonumber \\
& &\left.+2{\tilde H}^{{\mu}{\nu}}\left(F_{{\mu}{\nu}}-{\partial}_{[{\mu}}A_{{\nu}]}\right)\right\},
\end{eqnarray}
where ${\tilde H}^{{\mu}{\nu}}$ is introduced now as an auxiliary tensor field and $F_{{\mu}{\nu}}$ is regarded at present as an independent tensor field.
Now varying eq.~(3) with respect to ${\tilde H}^{{\mu}{\nu}}$ simply gives the definition of the abelian field strength eq.~(2),
together with which eq.~(3) turns back to the mother action
eq.~(1). This does not provide anything new but just shows the classical equivalence between the two
action forms. However, varying eq.~(3) with respect to $A_{\mu}$
leads to the equation that ${\tilde H}^{{\mu}{\nu}}$ satisfies,
\begin{equation}
{\partial}_{\mu}{\tilde H}^{{\mu}{\nu}}=0,
\end{equation}
which can be solved in terms of the Hodge dual of a $(p-1)$-form field strength ${\partial}_{[{\rho}}{\tilde A}_{{\sigma}_{1}{\cdots}{\sigma}_{p-2}]}$,
\begin{equation}
{\tilde H}^{{\mu}{\nu}}=\frac{1}{(p-1)!}{\epsilon}^{{\mu}{\nu}{\rho}{\sigma}_{1}{\cdots}{\sigma}_{p-2}}
{\partial}_{[{\rho}}{\tilde A}_{{\sigma}_{1}{\cdots}{\sigma}_{p-2}]},
\end{equation}
where ${\epsilon}^{{\mu}{\nu}{\rho}{\cdots}}$ is the alternating tensor density in $p+1$ spacetime dimensions and
${\tilde A}_{{\sigma}_{1}{\cdots}{\sigma}_{p-2}}$ is a $(p-2)$-form potential that is introduced for solving ${\tilde H}^{{\mu}{\nu}}$.
Next, dealing with $F_{{\mu}{\nu}}$ as an independent variable, we obtain
its field equation by making variation of eq.~(3) with respect to this tensor field,
\begin{equation}
-(-g)^{\frac{1}{4}}(-{\gamma})^{\frac{1}{4}}\left(g^{{\mu}{\rho}}F_{{\rho}{\sigma}}{\gamma}^{{\sigma}{\nu}}+{\gamma}^{{\mu}{\rho}}
F_{{\rho}{\sigma}}g^{{\sigma}{\nu}}
\right)=2{\tilde H}^{{\mu}{\nu}},
\end{equation}
where ${\tilde H}^{{\mu}{\nu}}$ is given at present by the solution eq.~(5).

In the present stage, the usual way of deriving the dual of eq.~(1), as adopted in ref.~\cite{s14}, is to solve from eq.~(6)
the tensor field $F_{{\mu}{\nu}}$
in terms of
${\gamma}_{{\mu}{\nu}}$, $g_{{\mu}{\nu}}$, and ${\tilde H}^{{\mu}{\nu}}$ and then to substitute the solution into eq.~(3).
Because of the complexity of eq.~(6), such a solution contains an infinite power series in the
auxiliary metric ${\gamma}_{{\mu}{\nu}}$. Nevertheless, the merit of the corresponding dual action is obvious, that is, this dual action is
quadratic in the new two-form field strength ${\tilde H}^{{\mu}{\nu}}$.

Instead of solving eq.~(6) directly, we provide an alternative way to deal with this equation. Considering the variation of eq.~(3)
with respect to ${\gamma}_{{\mu}{\nu}}$,
we have the field equation of the auxiliary metric
\begin{equation}
{\gamma}_{{\mu}{\nu}}=g_{{\mu}{\nu}}-
F_{{\mu}{\rho}}g^{{\rho}{\sigma}}F_{{\sigma}{\nu}},
\end{equation}
which takes the same form as that derived from eq.~(1), i.e., the addition of the Lagrange multiplier to eq.~(1) does not change the formulation of
the field equation of the auxiliary metric. Note that $F_{{\mu}{\nu}}$ of eq.~(3) should be treated as an implicit functional of
${\gamma}_{{\mu}{\nu}}$ that is now constrained
by eq.~(6) in the derivation of eq.~(7), while $F_{{\mu}{\nu}}$ of eq.~(1) is independent of the auxiliary metric.
It is easier to derive eq.~(7) from eq.~(1) as the two actions, eq.~(1) and eq.~(3), are equivalent classically.
Alternatively,
one derives from the action eq.~(3) the equations of motion firstly for ${\gamma}_{{\mu}{\nu}}$
and then for $F_{{\mu}{\nu}}$, which gives rise to the same formulations as eq.~(7) and eq.~(6), respectively. The reason is obvious, that is,
the equations of motion for $F_{{\mu}{\nu}}$ and ${\gamma}_{{\mu}{\nu}}$ are independent of
the order they are deduced from the same action.

Several corollaries of the field equation of the auxiliary metric eq.~(7) can be obtained, among which the useful one for our purpose takes the form
\begin{equation}
g^{{\mu}{\rho}}F_{{\rho}{\sigma}}{\gamma}^{{\sigma}{\nu}}={\gamma}^{{\mu}{\rho}}
F_{{\rho}{\sigma}}g^{{\sigma}{\nu}},
\end{equation}
see Appendix for its proof. With this relation, $F_{{\mu}{\nu}}$ can be solved easily from eq.~(6),
\begin{equation}
F_{{\mu}{\nu}}=-(-g)^{-\frac{1}{4}}(-{\gamma})^{-\frac{1}{4}}{\gamma}_{{\mu}{\rho}}{\tilde H}^{{\rho}{\sigma}}
g_{{\sigma}{\nu}},
\end{equation}
which, different from that given by ref.~\cite{s14}, is no longer an infinite power series.
Now substituting eq.~(9) into eq.~(3), we thus derive an alternative dual action
\begin{equation}
S_{\rm D}=-\frac{T_{p}}{4}\int d^{p+1}x(-g)^{\frac{1}{4}}(-{\gamma})^{\frac{1}{4}}\left\{\left[{\gamma}^{{\mu}{\nu}}g_{{\mu}{\nu}}
-\left(p-3\right)\right]
+(-g)^{-\frac{1}{2}}(-{\gamma})^{-\frac{1}{2}}{\tilde H}^{{\mu}{\sigma}}{\gamma}_{{\mu}{\rho}}g_{{\nu}{\sigma}}{\tilde H}^{{\nu}{\rho}}\right\}.
\end{equation}
Note that the new dual action does not involve any infinite power series and in particular the term related to ${\tilde H}^{{\mu}{\nu}}$ is,
just formally,
the first term of that
infinite power series~\cite{s14}.
In fact this expression reveals in physics an effective interaction that is realized by the price of the non-linearity in the new two-form field
strength ${\tilde H}^{{\mu}{\nu}}$
as emphasized above (see also eq.~(11) below).
The new dual action seems to be quadratic in ${\tilde H}^{{\mu}{\nu}}$, but in fact highly nonlinear.
The reason is that the auxiliary metric ${\gamma}_{{\mu}{\nu}}$ in eq.~(10) is no longer a free variable but an implicit functional of
$g_{{\mu}{\nu}}$ and ${\tilde H}^{{\mu}{\nu}}$ as follows:
\begin{equation}
{\gamma}_{{\mu}{\nu}}=g_{{\mu}{\nu}}-(-g)^{-\frac{1}{2}}(-{\gamma})^{-\frac{1}{2}}{\gamma}_{{\mu}{\rho}}{\tilde H}^{{\rho}{\sigma}}
{\gamma}_{{\sigma}{\lambda}}
{\tilde H}^{{\lambda}{\kappa}}g_{{\kappa}{\nu}},
\end{equation}
which is obtained when eq.~(9) is substituted into eq.~(7).
In principle, an equivalent dual action $S_{\rm D}[g_{{\mu}{\nu}},{\tilde H}^{{\mu}{\nu}}]$ can then be derived
by solving ${\gamma}_{{\mu}{\nu}}$ from eq.~(11) and substituting the solution into eq.~(10), but in practice this procedure is quite difficult
to carry out because of the non-linearity in ${\gamma}_{{\mu}{\nu}}$ in eq.~(11). Although the procedure is not performed both to
our result and to that of ref.~\cite{s14}, the explicit relation, i.e. eq.~(11) involved by the three variables
(${\gamma}_{{\mu}{\nu}}$, $g_{{\mu}{\nu}}$, and ${\tilde H}^{{\mu}{\nu}}$)
is provided here while such a relation
would be tedious to obtain from the action with the infinite power series in ${\gamma}_{{\mu}{\nu}}$ by meanings of a perturbative way
suggested in ref.~\cite{s14}.

Anyway, we give an alternative dual action with respect to U(1) gauge fields for the Born-Infeld theory and note that it is
characterized by a new interaction formulation. As to the motivation, we try to reveal as many dualities as possible existed in
both the BIons and $Dp$-branes
(to be discussed soon) now in this paper and already in ref.~\cite{s18} from a different point of view, i.e., we have actually done
for other interesting objects, such as chiral bosons and bosonic $p$-branes in the previous work~\cite{s21}.

We can write several different formulations of the dual action eq.~(10) which might be of interest. The first is related to the following definition
of $H^{{\mu}{\nu}}$,
\begin{equation}
H^{{\mu}{\nu}} \equiv (-g)^{-\frac{1}{4}}(-{\gamma})^{-\frac{1}{4}}{\tilde H}^{{\mu}{\nu}},
\end{equation}
with which the dual action takes a seemingly elegant form
\begin{equation}
S_{\rm D}=-\frac{T_{p}}{4}\int d^{p+1}x(-g)^{\frac{1}{4}}(-{\gamma})^{\frac{1}{4}}\left[g_{{\mu}{\nu}}\left({\gamma}^{{\mu}{\nu}}
+{\gamma}_{{\rho}{\sigma}}H^{{\mu}{\rho}}H^{{\sigma}{\nu}}\right)
-\left(p-3\right)\right].
\end{equation}
When comparing eq.~(13) with its mother form eq.~(1), we can see the so-called beauty of duality in formalism, that is, one action switches to the other
with the permutations of $g_{{\mu}{\nu}}$ and ${\gamma}^{{\mu}{\nu}}$, of
$H^{{\mu}{\nu}}$ and $F_{{\mu}{\nu}}$, and of plus sign and minus sign in the first term of the square bracket,
which presents a quite interesting symmetry.
However, we have to note that
${\gamma}_{{\mu}{\nu}}$ is not free but constrained by
eq.~(11) and the divergence of $H^{{\mu}{\nu}}$, different from that of ${\tilde H}^{{\mu}{\nu}}$, is not equal to zero identically in eq.~(13).
In addition, if we redefine the spacetime metric and its corresponding auxiliary one as follows:
\begin{equation}
{\tilde g}_{{\mu}{\nu}}\equiv \frac{g_{{\mu}{\nu}}}{\sqrt{-g}}, \hspace{20pt}
{\tilde{\gamma}}_{{\mu}{\nu}}\equiv \frac{{\gamma}_{{\mu}{\nu}}}{\sqrt{-{\gamma}}},
\end{equation}
we therefore turn to another different formulation of the dual action
\begin{eqnarray}
S_{\rm D}&=&-\frac{T_{p}}{4}\int d^{p+1}x(-{\tilde g})^{-\frac{1}{2(p-1)}}(-{\tilde{\gamma}})^{-\frac{1}{2(p-1)}}
\left\{\left[(-{\tilde{\gamma}})^{\frac{1}{p-1}}(-{\tilde g})^{-\frac{1}{p-1}}{\tilde{\gamma}}^{{\mu}{\nu}}{\tilde g}_{{\mu}{\nu}}
-\left(p-3\right)\right]\right.\nonumber \\
& &\left.+{\tilde H}^{{\mu}{\sigma}}{\tilde{\gamma}}_{{\mu}{\rho}}{\tilde g}_{{\nu}{\sigma}}{\tilde H}^{{\nu}{\rho}}\right\}.
\end{eqnarray}
It looks a little bit of complexity with the powers of the determinants, but the term related to the field strength ${\tilde H}^{{\mu}{\nu}}$
seems to be simpler.
In particular, its four-dimensional case
\begin{equation}
S_{\rm D}(p=3)=-\frac{T_{3}}{4}\int d^{4}x(-{\tilde g})^{-\frac{1}{4}}(-{\tilde{\gamma}})^{-\frac{1}{4}}
\left(\sqrt{-{\tilde{\gamma}}}{\tilde{\gamma}}^{{\mu}{\nu}}\frac{{\tilde g}_{{\mu}{\nu}}}{\sqrt{-{\tilde g}}}
+{\tilde H}^{{\mu}{\sigma}}{\tilde{\gamma}}_{{\mu}{\rho}}{\tilde g}_{{\nu}{\sigma}}{\tilde H}^{{\nu}{\rho}}\right),
\end{equation}
shows its conformal invariance obviously under the Weyl transformation of the new auxiliary metric
\begin{equation}
{\tilde{\gamma}}_{{\mu}{\nu}} \longrightarrow {\omega}{\tilde{\gamma}}_{{\mu}{\nu}},
\end{equation}
where ${\omega}$ is an arbitrary real function of spacetime coordinates.

\section{Dual Actions for the $Dp$-branes}

As the $Dp$-brane kinetic term takes the form of Born-Infeld type~\cite{s3},
the dualization is therefore quite similar.
The starting point is the action~\cite{s14}
\begin{equation}
S=-\frac{T_{p}}{4}\int d^{p+1}{\xi}e^{-\phi}(-g)^{\frac{1}{4}}(-{\gamma})^{\frac{1}{4}}
\left[{\gamma}^{{\mu}{\nu}}\left(g_{{\mu}{\nu}}-g^{{\rho}{\sigma}}{\cal F}_{{\mu}{\rho}}{\cal F}_{{\sigma}{\nu}}\right)-(p-3)
\right],
\end{equation}
where
\begin{equation}
{\cal F}_{{\mu}{\nu}}\equiv F_{{\mu}{\nu}}-B_{{\mu}{\nu}},
\end{equation}
$\phi$, $g_{{\mu}{\nu}}$ and $B_{{\mu}{\nu}}$ are pullbacks to the worldvolume of the background dilaton, spacetime metric
and NS antisymmetric two-form fields,
and $F_{{\mu}{\nu}}={\partial}_{\mu}A_{\nu}-{\partial}_{\nu}A_{\mu}$, with $A_{\mu}({\xi})$ the $U(1)$ worldvolume gauge field.
${\gamma}_{{\mu}{\nu}}$ is the auxiliary worldvolume metric that is introduced for the same purpose as that of
the Born-Infeld case. Now Greek lowercase letters, ${\mu},{\nu},{\rho},\cdots$, running over $0,1,\cdots,p$,
are utilized as indices in the worldvolume
that is spanned by $p+1$ arbitrary parameters ${\xi}^{\mu}$. The next step is to add a Lagrange multiplier term to the above action and
therefore one
obtains its classically equivalent form
\begin{eqnarray}
S^{\prime}&=&-\frac{T_{p}}{4}\int d^{p+1}{\xi}\left\{e^{-\phi}(-g)^{\frac{1}{4}}(-{\gamma})^{\frac{1}{4}}
\left[{\gamma}^{{\mu}{\nu}}\left(g_{{\mu}{\nu}}-g^{{\rho}{\sigma}}{\cal F}_{{\mu}{\rho}}{\cal F}_{{\sigma}{\nu}}\right)-(p-3)\right]
\right.\nonumber \\
& &\left.+2{\tilde H}^{{\mu}{\nu}}\left(F_{{\mu}{\nu}}-{\partial}_{[{\mu}}A_{{\nu}]}\right)\right\}.
\end{eqnarray}
Making variation of eq.~(20) with respect to $A_{\mu}$ leads to the same expression as eq.~(5) in which ${\tilde H}^{{\mu}{\nu}}$ has been solved
in terms of a
$(p-2)$-form potential ${\tilde A}_{{\sigma}_{1}{\cdots}{\sigma}_{p-2}}$ in the worldvolume. Moreover, doing for the two-form
$F_{{\mu}{\nu}}$ gives its field equation
\begin{equation}
-e^{-\phi}(-g)^{\frac{1}{4}}(-{\gamma})^{\frac{1}{4}}\left(g^{{\mu}{\rho}}{\cal F}_{{\rho}{\sigma}}{\gamma}^{{\sigma}{\nu}}+{\gamma}^{{\mu}{\rho}}
{\cal F}_{{\rho}{\sigma}}g^{{\sigma}{\nu}}
\right)=2{\tilde H}^{{\mu}{\nu}},
\end{equation}
which looks like eq.~(6) formally just with the replacement of $F_{{\mu}{\nu}}$ by $e^{-\phi}{\cal F}_{{\mu}{\nu}}$.
The following step is to write
the field equation of the auxiliary metric derived from either eq.~(18) or eq.~(20),
\begin{equation}
{\gamma}_{{\mu}{\nu}}=g_{{\mu}{\nu}}-{\cal F}_{{\mu}{\rho}}g^{{\rho}{\sigma}}{\cal F}_{{\sigma}{\nu}},
\end{equation}
which, through a similar proof to that of the Born-Infeld theory, also leads to a useful corollary
\begin{equation}
g^{{\mu}{\rho}}{\cal F}_{{\rho}{\sigma}}{\gamma}^{{\sigma}{\nu}}={\gamma}^{{\mu}{\rho}}
{\cal F}_{{\rho}{\sigma}}g^{{\sigma}{\nu}}.
\end{equation}
Using this relation, one can solve ${\cal F}_{{\mu}{\nu}}$ easily from eq.~(21),
\begin{equation}
{\cal F}_{{\mu}{\nu}}=-e^{\phi}(-g)^{-\frac{1}{4}}(-{\gamma})^{-\frac{1}{4}}{\gamma}_{{\mu}{\rho}}{\tilde H}^{{\rho}{\sigma}}
g_{{\sigma}{\nu}}.
\end{equation}
The final step is thus to substitute eq.~(24) into eq.~(20) and at last one deduces an alternative dual action for the $Dp$-branes,
\begin{eqnarray}
S_{\rm D}&=&-\frac{T_{p}}{4}\int d^{p+1}{\xi}e^{-\phi}(-g)^{\frac{1}{4}}(-{\gamma})^{\frac{1}{4}}\left\{\left[{\gamma}^{{\mu}{\nu}}g_{{\mu}{\nu}}
-\left(p-3\right)\right]
+2e^{\phi}(-g)^{-\frac{1}{4}}(-{\gamma})^{-\frac{1}{4}}{\tilde H}^{{\mu}{\nu}}B_{{\mu}{\nu}}\right.\nonumber \\
& &\left.+e^{2\phi}(-g)^{-\frac{1}{2}}(-{\gamma})^{-\frac{1}{2}}{\tilde H}^{{\mu}{\sigma}}{\gamma}_{{\mu}{\rho}}g_{{\nu}{\sigma}}
{\tilde H}^{{\nu}{\rho}}\right\}.
\end{eqnarray}
One can see that this kind of dual actions has nothing to do with any infinite power series in ${\gamma}_{{\mu}{\nu}}$,
and that the term related to the square of ${\tilde H}^{{\mu}{\nu}}$ is just the first term of that infinite power series appeared in ref.~\cite{s14}.
The field equation of the auxiliary metric turns out to be the constrained condition
\begin{equation}
{\gamma}_{{\mu}{\nu}}=g_{{\mu}{\nu}}-e^{2\phi}(-g)^{-\frac{1}{2}}(-{\gamma})^{-\frac{1}{2}}{\gamma}_{{\mu}{\rho}}{\tilde H}^{{\rho}{\sigma}}
{\gamma}_{{\sigma}{\lambda}}
{\tilde H}^{{\lambda}{\kappa}}g_{{\kappa}{\nu}}.
\end{equation}
Eq.~(25), together with eq.~(26), shows a kind of effective interactions involving finite terms in physics.
It is the non-linearity in ${\tilde H}^{{\mu}{\nu}}$ that this action is
classically equivalent to that of ref.~\cite{s14}.
Incidentally, when ${\tilde H}^{{\mu}{\nu}}$ is replaced by $H^{{\mu}{\nu}}$ defined by
\begin{equation}
H^{{\mu}{\nu}} \equiv e^{\phi}(-g)^{-\frac{1}{4}}(-{\gamma})^{-\frac{1}{4}}{\tilde H}^{{\mu}{\nu}},
\end{equation}
the dual action eq.~(25) turns out to be much simpler in formalism,
\begin{equation}
S_{\rm D}=-\frac{T_{p}}{4}\int d^{p+1}{\xi}e^{-\phi}(-g)^{\frac{1}{4}}(-{\gamma})^{\frac{1}{4}}
\left[g_{{\mu}{\nu}}\left({\gamma}^{{\mu}{\nu}}
+{\gamma}_{{\rho}{\sigma}}H^{{\mu}{\rho}}H^{{\sigma}{\nu}}\right)
+2H^{{\mu}{\nu}}B_{{\mu}{\nu}}-\left(p-3\right)\right].
\end{equation}
The further discussions on the dual actions of $Dp$-branes are similar to that of the Born-Infeld and omitted here.

\vspace{10mm}
\noindent
{\bf Acknowledgments}

\vspace{2mm}
\noindent
This work was supported in part by the National Natural
Science Foundation of China under grant No.10675061 and by the Ministry of Education of China under grant No.20060055006.
Y.-G. Miao would like to thank the Abdus Salam International Centre for
Theoretical Physics for hospitality where part of the work was performed.

\newpage
\section*{Appendix  Proof of the Formula $G^{-1}F{\Gamma}^{-1}={\Gamma}^{-1}FG^{-1}$}
Here $G$, ${\Gamma}$, and $F$ stand for the matrices of $g_{{\mu}{\nu}}$, ${\gamma}_{{\mu}{\nu}}$, and $F_{{\mu}{\nu}}$, respectively.
As a result, the above formula is the matrix form of the component one
$g^{{\mu}{\rho}}F_{{\rho}{\sigma}}{\gamma}^{{\sigma}{\nu}}={\gamma}^{{\mu}{\rho}}F_{{\rho}{\sigma}}g^{{\sigma}{\nu}}$.
The starting point of this proof is the field equation of the auxiliary metric eq.~(7) whose matrix form can be written as ${\Gamma}=G-FG^{-1}F$.

{\em Lemma}: If two matrices $A$ and $B$ are commutative, i.e. $AB=BA$, and one of them is invertible, say $A$,
then $A^{-1}$ and $B$ must be commutative, i.e. $A^{-1}B=BA^{-1}$.

The proof of this lemma is obvious, that is, multiplying by $A^{-1}$ to the left and right successively on both sides of $AB=BA$
leads directly to $A^{-1}B=BA^{-1}$.

Let us turn to the proof of the formula. Rewrite the field equation of the auxiliary metric as
\begin{equation}
{\Gamma}=G(1-G^{-1}FG^{-1}F),
\end{equation}
whose inverse thus takes the form
\begin{equation}
{\Gamma}^{-1}=(1-G^{-1}FG^{-1}F)^{-1}G^{-1}.
\end{equation}
Multiplying by $F$ to the right on both sides of the above equation gives
\begin{equation}
{\Gamma}^{-1}F=(1-G^{-1}FG^{-1}F)^{-1}G^{-1}F.
\end{equation}
If let $A\equiv 1-G^{-1}FG^{-1}F$ and $B\equiv G^{-1}F$, with the lemma eq.~(31) turns out to be
\begin{eqnarray}
{\Gamma}^{-1}F&=&G^{-1}F(1-G^{-1}FG^{-1}F)^{-1}\nonumber \\
&=&G^{-1}F{\Gamma}^{-1}G,
\end{eqnarray}
where $(1-G^{-1}FG^{-1}F)^{-1}={\Gamma}^{-1}G$ has been used to the last equality. Eq.~(32) is just the formula after $G^{-1}$ is multiplied
to the right on both sides of the equation.

\newpage

\end{document}